\input ./style/arxiv-general.cfg
\documentclass[aos,MSNbibl,seceqn,dvips]{arximspdf}
\makeatletter
   \@ifpackageloaded{graphicx}{}{\usepackage{graphicx}}
\makeatother
\usepackage{algorithm}
\usepackage{algpseudocode}

\doi{10.1214/15-AOS1338}
\volume{43}
\issue{6}
\pubyear{2015}
\firstpage{2384}
\lastpage{2411}
\docsubty{FLA}

\makeatletter
\renewcommand{\algorithmicrequire}{\textbf{Input:}}
\renewcommand{\algorithmicensure}{\textbf{Output:}}

\newcommand{\eqref}[1]{(\ref{#1})}
\newtheorem{theorem}{Theorem}
\newtheorem{lemma}[theorem]{Lemma}
\newtheorem{proposition}[theorem]{Proposition}
\newproclaim{definition}[theorem]{Definition}
\newproclaim{example}[theorem]{Example}

\def\E{{\mathbb E}}
\def\var{\operatorname{Var}}
\def\cov{\operatorname{Cov}}
\renewcommand{\Pr}{ \mathbb P}
\newcommand{ \s}{ \mathcal S }
\newcommand{ \cL}{ \mathcal L }
\newcommand{ \cV}{ \mathcal V}
\newcommand{\ga}{\alpha}
\newcommand{\gve}{\varepsilon}
\newcommand{\gt}{\theta}
\newcommand{\gl}{\lambda}
\newcommand{\gs}{\sigma}
\newcommand{\rar}{ \rightarrow}
\newcommand{\til}{\tilde}
\newcommand{\V}[1]{\bolds{#1}}
\makeatother

\begin{document}
\begin{frontmatter}

\title{Subsampling bootstrap of count features of~networks\thanksref{T2}}
\runtitle{Subsampling bootstrap of networks}

\begin{aug}
\author[A]{\fnms{Sharmodeep}~\snm{Bhattacharyya}\corref{}\thanksref{m1,m2}\ead[label=e1]{bhattash@science.oregonstate.edu}}
\and
\author[B]{\fnms{Peter J.}~\snm{Bickel}\thanksref{m2}\ead[label=e2]{bickel@stat.berkeley.edu}}
\runauthor{S. Bhattacharyya and P.~J. Bickel}
\thankstext{T2}{Supported in part by NSF Grant DMS-11-60319.}
\affiliation{University of California, Berkeley\thanksmark{m1} and Oregon
State University\thanksmark{m2}}
\address[A]{Department of Statistics\\
Oregon State University\\
44 Kidder Hall\\
Corvallis, Oregon 97331\\
USA\\
\printead{e1}}
\address[B]{Department of Statistics\\
University of California, Berkeley\\
367 Evans Hall\\
Berkeley, California 94720\\
USA\\
\printead{e2}}
\end{aug}

%
\received{\smonth{2} \syear{2014}}
%
\revised{\smonth{4} \syear{2015}}

%
\begin{abstract}
Analysis of stochastic models of networks is quite important in
light of the huge influx of network data in social, information and bio
sciences, but a proper statistical analysis of features of different
stochastic models of networks is still underway. We propose bootstrap
subsampling methods for finding empirical distribution of count
features or ``moments'' (Bickel, Chen and Levina [\textit{Ann.
Statist.} \textbf{39} (2011) 2280--2301]) and smooth functions of these features
for the networks. Using these methods, we cannot only estimate the
variance of count features but also get good estimates of such feature
counts, which are usually expensive to compute numerically in large
networks. In our paper, we prove theoretical properties of the
bootstrap estimates of variance of the count features as well as show
their efficacy through simulation. We also use the method on some real
network data for estimation of variance and expectation of some count features.
\end{abstract}

%

%
\end{frontmatter}

\section{Introduction}
\label{intro}
The study of networks has received recent increased attention, not only
in social sciences, mathematics and statistics, but also in physics and
computer science. With the information boom, a huge number of network
data sets have appeared. In biology, gene regulation networks,
protein--protein interaction networks, neural networks, ecological and
epidemiological networks have become increasingly important. In social
media, the Facebook, Twitter and Linkedin networks have come into
prominence. Information networks have arisen in connection with text
mining. Technological networks such as the Internet and many other
networks related to Internet have also become objects of study.

In this paper, we consider a nonparametric formulation for network
models where node labels carry no information. The model was proposed
in Bickel and Chen \cite{bickel2009nonparametric} and has its origins
in the works of Aldous \cite{MR637937} and Hoover \cite{hoover1979relations}.
Exchangeable probability models on infinite networks have a general
representation based on the results of Aldous \cite{MR637937}, Hoover
\cite{hoover1979relations}, Kallenberg \cite{MR2161313} and Diaconis
and Janson \cite{MR2463439}. The result is analogous to de Finetti's
theorem. Note that numerical representation of networks come in the
form of the adjacency matrix $A$, where $A_{ij} = 1$ if there is an
edge from node $i$ to $j$ and $0$ otherwise. We assume $A_{ii} = 0$;
that is, there are no self-loops. It is natural to assume exchangeable
property for probability distribution on unlabeled random networks,
which means that the probability distribution on the set of all
adjacency matrices $\cL\{[Aij ], i, j \geq1\}$ satisfy $\cL\{
[A_{ij}]\} = \cL\{[A_{\gs_i \gs_j} ]\}$, where $\gs$ is an arbitrary
permutation function on node indices. Such exchangeable probability
distributions on random infinite binary arrays can be characterized as
follows: for $i,j \geq1$,
\begin{eqnarray*}
\ga, \xi_i, \eta_{ij} & \stackrel{\mathrm{i.i.d.}} {\sim}& U(0,1),
\\
A_{ij} & = & f(\ga, \xi_i, \xi_j,
\eta_{ij}),
\end{eqnarray*}
where, $f:[0,1]^4 \rar[0,1]$ is a measurable function, symmetric in
its second and third arguments and $\eta_{ij}=\eta_{ji}$. $\ga$, as in
de Finetti's theorem, corresponds to the mixing distribution and is not
identifiable. This representation is not unique, and $f$ is not
identifiable. These distributions can be parametrized through the function
\[
h(u,v)=P [A_{ij} =1| \xi_i = u, \xi_j =v ].
\]
The function $h$ is still not unique, but it can be shown that if two
functions $h_1$ and $h_2$ define the same distribution $\cL$, they can
be related through a measure-preserving transformation. This leads to
the Bickel and Chen \cite{bickel2009nonparametric} characterization of
``nonparametric'' unlabeled graph models, which is closely related to
Lov\'{a}sz's notion of ``graphons'' \cite{lovasz2012large}. The model
will be described in more detail in Section~\ref{main}. Other
researchers have also studied similar, general classes of models, such
as the \textit{latent space models} of Hoff, Raftery and Handcock
\cite
{MR1951262} and the \emph{inhomogeneous random graph models} of
Bollob\'
{a}s, Janson and Riordan \cite{MR2337396}. Many previously studied
probability models for networks fall into this class. The class
includes the stochastic block models (Holland, Laskey and Leinhardt
\cite{holland1983stochastic}, Nowicki and Snijders \cite{MR1947255})
and the configuration model (Chung and Lu \cite{MR1955514}).
Dynamically defined models such as the ``preferential attachment''
models (which seem to have been first mentioned by Yule in the 1920s
and given its modern name by Barab{\'a}si and Albert \cite
{barabasi1999emergence}) can also be thought of in this way if the
dynamical construction process continues forever, producing an infinite
graph. More details are given in Section~\ref{simulation}.

Motifs or count statistics are the main statistics that we consider in
this paper. Count statistics can be defined as smooth functions of
counts of subgraphs in the network. Counts of special subgraphs have
been extensively used in the network literature for analyzing network
behavior \cite{bearman2004chains,milo2004superfamilies,przytycka2006important}. The count statistics have appeared earlier
under the names \textit{motif} counts in biology \cite{milo2002network}
and \textit{subgraph} counts in probability \cite{lovasz2012large}. It
also follows from the work of Lov\'{a}sz \cite{lovasz2012large},
Diaconis and Janson \cite{MR2463439} and in part from Bickel and Chen
\cite{bickel2009nonparametric} that there is a unique set of statistics
whose joint distribution characterize the probability distribution on
unlabeled networks. These statistics, called \textit{empirical
moments} by
Bickel, Chen and Levina \cite{MR2906868} are the counts of subgraphs in
the network. The subgraphs most used are small cycles like triad,
tetrad and small acyclic graphs.

The expectation and variances of count statistics can, in principle, be
computed (Picard et al. \cite{picard2008assessing}) and more usefully
be asymptotically approximated \cite{MR2906868}. Under appropriate
conditions, normalized count statistics have limiting Gaussian
distribution. They have many uses \cite{wasserman1994social,MR2834086,bearman2004chains}, particularly in distinguishing
between the mechanisms generating different graphs as well as providing
characterization of network distributions. The general asymptotic
Gaussian distribution of count statistics was provided in Theorem~1 of
\cite{MR2906868} with an expression for the asymptotic mean and
variance; however, the paper provided no way to calculate the quantities.

Motifs or count statistics have been used in testing equality of
features of networks and finding confidence intervals of the count
features \cite{shen2002network,middendorf2005inferring}.
However, a major stumbling block in their use has been the calculation
of motifs that have even moderately large number of vertices (i.e.,
more than five) and even more challenging problem of finding
estimates of their variances. Finding the correct count statistics or
motifs is a computationally hard problem for large networks, as the
complexity of finding the count of a subgraph is polynomial in terms of
number of vertices, and when the number of vertices in the network is
even in thousands, the computation becomes difficult; if it is in
millions, the computation becomes infeasible.
Using subsampling methods to calculate the count statistics, we can
greatly reduce the computational burden of computing the statistics and
inference using them.

In the statistical literature on networks, some work has been done on
devising sampling designs to select network samples. Various sampling
designs have been proposed in the statistical and computer science
literature to derive \emph{representative} samples of a given network;
see \cite{MR2724362,leskovec2005graphs} and \cite{MR2894042}.
Many of these sampling designs have been analyzed from the design-based
sampling point of view \cite{thompson2000model,frank2005network}. Some of these methods have been analyzed from a
model-based sampling point of view, where mostly the \emph{exponential
random graph model (ERGM)} has been considered as the model generating
the network, and a likelihood-based approach has been taken for
inference \cite{handcock2010modeling}. As a result, only parametric
inference was possible. On the other hand, our approach is not
restricted to parametric models as we try to estimate the certain
functionals of the underlying nonparametric generating model, using the
samples obtained from the network data.

\subsection{Contribution and structure of our work}
We use subsampling-based bootstrap approaches to estimate the count
statistics as well as find the approximate distributions for such count
statistics under the general model of Bickel and Chen~\cite
{bickel2009nonparametric}.


Along with the bootstrap methods and their theoretical analysis, we
give two examples where the use of count statistics provides some
useful insights into the behavior of the networks. One of the two
examples is the Jefferson High School network given in Bearman et al.
\cite{bearman2004chains}, and the other example uses the Facebook
collegiate networks provided in Traud et al. \cite{MR2724362}.
The high school network is a nice example where counts of specific
types of subgraphs in the network and their confidence intervals based
on different generating models give us useful insight into the behavior
of nodes in the network \cite{bearman2004chains}.
The Facebook collegiate networks are larger and denser networks, and
calculation of count statistics for these networks would not be
computationally feasible without the use of subsampling methods.

In Section~\ref{main} we outline our main results. In Section~\ref
{methods} we describe the bootstrap subsampling methods and the
theoretical properties of each bootstrap estimator. We also indicate a
method for estimating asymptotic variances of these estimators using
bootstrap. Additionally, we give a theoretical comparison of the methods.
In Section~\ref{sec_theory}, we give the general theorem on asymptotic
Gaussianity of bootstrap subsampling estimates count statistics and
their variance. In Section~\ref{simulation} we perform a simulation
study under two special cases of the general ``nonparametric'' model:
the stochastic block model and the preferential attachment model, respectively.
In Section~\ref{real data} we apply our method to test hypotheses about
the count statistics of real networks.

\section{Model and statistics}
\label{main}
We consider a random unlabeled graph $G_n$ as the data. Let $V(G_n) = \{
v_i, \ldots, v_n\}$ denote the vertices of $G_n$ and $E(G_n)$ denote
the set of edges of $G_n$. Thus the number of vertices in $G_n$ is
$|V(G_n)| = n$.
We shall only consider undirected, unweighted graphs in this paper.
For the sake of notational simplicity, we may denote $G_n$ by $G$.

As usual we suppose the network is represented by an adjacency matrix
$A_{n\times n}$ whose elements are $A_{ij}\in\{0,1\}$,
\[
A_{ij} = %
\cases{ 1, & \quad $\mbox{if node $i$ links to node $j$,}
$\vspace*{2pt}
\cr
0, & \quad $\mbox{otherwise.}$} %
\]
%


A finite sample version of the Aldous--Hoover representation for
exchangeable adjacency matrices $A_{n\times n}$ becomes, for $i,j \in
\{1,\ldots,n\}$,
\begin{eqnarray*}
\ga, \xi_i, \eta_{ij} & \stackrel{\mathrm{i.i.d.}} {\sim}& U(0,1),
\\
A_{ij} & = & f_n(\ga,\xi_i,
\xi_j, \eta_{ij}),
\end{eqnarray*}
where, $f_n:[0,1]^4 \rar[0,1]$ is a measurable function, symmetric in
its second and third arguments and $\eta_{ij}=\eta_{ji}$. Note that
this is not a representation of all exchangeable probability
distributions on finite networks.

Bickel and Chen \cite{bickel2009nonparametric} considered a special
form of the general Aldous--Hoover representation,
\[
h_n(u,v) \equiv\Pr(A_{ij}=1 | \xi_i=u,
\xi_j = v).
\]
%
The above-mentioned form can be simplified by decoupling $n$ from the
contribution of $(\xi_i, \xi_j)$. Thus $h_n$ is modeled as product of a
scale function in terms of $n$, $\rho_n$, defined as
\[
\rho_n = \int_0^1\int
_0^1h_n(u,v)\,du\,dv
\]
and a bivariate function independent of $n$, the \emph{latent variable
density}, $w(\xi_i,\xi_j)$. We call the resulting model a \emph
{nonparametric latent variable model}, and the model equation described
in Bickel, Chen and Levina \cite{MR2906868} becomes, for $i,j \in\{
1,\ldots,n\}$, $\xi_i \stackrel{\mathrm{i.i.d.}}{\sim} U(0,1)$ and
%
\begin{equation}
\label{eq:nonpar_model} \Pr(A_{ij} = 1 | \xi_i = u,
\xi_j = v) = h_n(u, v) = \rho_n w(u, v)
\mathbf{1}\bigl(w \leq\rho_n^{-1}\bigr),
\end{equation}
where $w(u, v) \geq0$, symmetric, $0\leq u, v\leq1$, $\int\int
w(u,v)\,du\,dv=1$, $0 <\rho_n < 1$ and we define expected degree $\gl_n =
n\rho_n$.

The graph statistics that we are concerned with are count statistics of
subgraphs. Let $R$ be a subgraph of $G$, with $V(R)\subseteq V(G)$ and
$E(R)\subseteq E(G)$. We have $|V(R)| = p$ and $|E(R)| = e$. For
notation, if two graphs $R$ and $S$ are equivalent, we denote them by
$R\cong S$, and if $R$ is a subgraph of $S$, we denote them by
$R\subseteq S$. The integral parameter corresponding to a subgraph $R$
is defined as $P(R)$,
%
\begin{equation}
\label{eq_int_func} P(R) = \E \biggl\{\prod_{(i,j)\in E(R)} h(
\xi_i, \xi_j)\prod_{(i,j)\in
E(\bar{R})}
\bigl(1 - h(\xi_i, \xi_j) \bigr) \biggr\},
\end{equation}
where $\bar{R}$ is a subgraph of $K_p$ ($K_p$ is a complete graph on
$p$ vertices) with $V(\bar{R}) = \{i,j: (i,j)\notin E(R), i,j \in
V(R)\}
$ and $E(\bar{R}) = \{(i,j): (i,j) \notin E(R), i\in V(R), j\in V(R)\}$.

Now, the empirical statistic corresponding to $P(R)$, which is the
count statistics for subgraph $R$, is
%
\begin{equation}
\label{statistic} \hat{P}(R) = \frac{1}{{n \choose p} |\operatorname{Iso}(R)|}\sum_{S\subseteq K_n,
S\cong R}
\mathbf{1}(S \subseteq G),
\end{equation}
where $\operatorname{Iso}(R)$ is the group of isomorphisms of $R$, and $K_n$ is the
complete graph on $n$ vertices.

We also have from \cite{MR2906868}
\[
\E\bigl(\hat{P}(R)\bigr) = P(R).
\]

Examples of subgraphs and corresponding count statistics include the following:

\begin{example}
\label{ex_edge}
$R = $ edge is a subgraph with two vertices and one edge connecting
them, so $\hat{P}(R) = \frac{1}{n(n-1)}\sum_{i=1}^n D_i$, where $D_i =
\mbox{degree of } v_i$, $v_i \in V(G)$. $P(R) = \int_0^1 h_n(u,v)\,du\,dv$.
\end{example}

\begin{example}
\label{ex_triangle}
$R = $ triangle is a 3-clique subgraph, so $\hat{P}(R) = \frac{1}{{n
\choose3}}$ total number  of unique 3-clique subgraphs in $G_n$,
\[
P(R) = \int_0^1\int_0^1\int_0^1 h_n(u,v)h_n(v,s) h_n(s,u)\,du\,dv\,ds.
\]
\end{example}

\begin{example}
\label{ex_transitivity}
We define a smooth function of counts of triangles and ``V's,'' known
as \textit{transitivity}, $T_{\mathrm{Tr}}$, as
\[
\hat{P}_{\mathrm{Tr}} = \frac{\hat{\rho}_n^{-3}\hat{P}(R_1)}{\hat{\rho
}_n^{-3}\hat{P}(R_1) + \hat{\rho}_n^{-2}\hat{P}(R_2)},
\]
where $R_1$ is a triangle or a 3-cycle, and $R_2$ is a ``V'' or a path
with three vertices and $\hat{\rho}_n = \hat{P}(\mbox{edge})$.
\end{example}

\begin{example}
\label{ex_cycle}
$R = p$-cycle is a cyclic subgraph with $|V(R)|=p$, $|E(R)|=p$, and $R$
is a \emph{ring} containing all $p$ vertices. Triangle is a $3$-cycle.
$P(R) = \int_0^1\cdots\int_0^1 h_n(u_1,u_2)\cdots
h_n(u_{p-1},u_p)h_n(u_{p},u_1)\,du_1\cdots \,du_p$.
\end{example}

\begin{definition}[(Wheels)]
A $(k, l)$-wheel is an acyclic graph with $kl + 1$ vertices and $kl$
edges and ``hub'' vertex (say, $\{1\}$), isomorphic to the graph with
edges $\{((1,2), (2,3),\ldots,(k,k + 1)) ((1,k + 2), (k+2, k+3),
\ldots, (2k, 2k + 1)),\ldots, ((1, (l-1)k + 2), ((l-1)k+2,
(l-1)k+3), \ldots, (lk, lk + 1))\}$.
\end{definition}

Edges, ``V,'' ``W'' are examples of $(k, l)$-wheels. An edge is a $(1,
1)$-wheel, a ``V'' is a $(1, 2)$-wheel and a ``W'' is a $(2, 2)$-wheel.

Now, as $\rho_n \rar0$, which is the case for graphs which are \emph
{not} fully dense, that is, $|E(G_n)| = O_P(n^2)$, $P(R) \rar0$ as
well as its estimator $\hat{P}(R) \stackrel{P}{\rar} 0$ and the
asymptotics on $(\hat{P}(R) - P(R))$ become uninformative. So, in order
to get a proper analysis of the behavior of $\hat{P}(R)$ in relation to
$P(R)$, we have to appropriately normalize both $P(R)$ and $\hat
{P}(R)$. The normalized versions of parameter $P(R)$ are defined as
%
\begin{equation}
\label{eq_norm_par} \tilde{P}(R) = \rho_n^{-e} P(R),
\end{equation}
where $e \equiv|E(R)|$. Then we define the corresponding normalized
statistic to be
%
\begin{equation}
\label{eq_norm_statistic} \hat{T}(R) = \hat{\rho}^{-e} \hat{P}(R),
\end{equation}
where
%
\begin{equation}
\label{eq_rho_def} \hat{\rho} = \frac{\bar{D}}{n - 1},
\end{equation}
where $D_i = \mbox{degree of } v_i$, $v_i\in V(G_n)$ for $i = 1,
\ldots
, n$ and $\bar{D} = \frac{1}{n}\sum_{i=1}^n D_i$. Now the investigation
on asymptotic behavior of $\sqrt{n}(\hat{T}(R) - \til{P}(R))$ is
possible, as both terms are asymptotically nonzero quantities. This
investigation was done in \cite{MR2906868}.

We wish to approximate the statistic $\hat{P}(R)$ and functional $\var
(\hat{P}(R))$ by nonparametric bootstrap. We consider two bootstrap procedures:
\begin{longlist}[(II)]
\item[(I)] the \textit{uniform subsampling} bootstrap procedure and
%
\item[(II)] the \textit{subgraph subsampling} bootstrap procedure.
\end{longlist}

How we get the bootstrap estimates will be discussed in next section,
and we will state theorems justifying the use of these bootstrap
estimations in next two sections.

\subsection{Bootstrap and model-based sampling}
Our work can be viewed from two different perspectives. The first
perspective is that of the \emph{bootstrap}. In \emph{nonparametric
bootstrap}, we use resamples or subsamples of the data, where the data
comes from an unknown distribution, to find the functionals of the
unknown distribution. In our situation also, we have a network that has
been generated from an underlying probability model. We want to \emph
{subsample} networks from our given network and use those subsampled
networks to approximate estimates of functionals of the underlying
population model generating the given network. Note that here we are
interested in the \emph{subsampling}, not the \emph{resampling} of a
network. Our use of the bootstrap corresponds to Efron's \cite
{MR515681} use of bootstrap for approximations made by Monte Carlo
quantities, which, in principle, could be calculated using data such as
the empirical variance of complicated estimates.

The second perspective is that of \emph{sampling}. In sampling, we
consider that the \emph{population}, from which the sample is selected
according to some sampling design, is a realization of a probabilistic
event. So, in our case, we consider the given network as the
population, and it is generated from an underlying probability model.
We use subsampling bootstrap or sampling of network data to get
estimates for population quantity (count statistics) and underlying
probability model (integral parameter).

\section{Bootstrap methods}
\label{methods}

We consider two different bootstrap methods. Both of the methods of
bootstrap consider finding subsamples from the whole network given as
the data. In the following subsections, we shall define each of these
subsampling bootstrap methods. We shall also compare the theoretical
performance between the two bootstrap schemes.

Let the adjacency matrix of $G_n$ be denoted by $A_{n\times n}$.
Let $R$ be a subgraph of $G$, with $V(R)\subseteq V(G)$ and
$E(R)\subseteq E(G)$. We have $|V(R)| = p$ and $|E(R)| = e$.

\subsection{Uniform subsampling bootstrap}
\label{naive}
In the \emph{uniform subsampling} bootstrap scheme, at each bootstrap
iteration, a subset of vertices of the full network $G$ is selected
without replacement, and the graph induced by the selected subset of
vertices is the subnetwork we consider. This is a vertex subsampling or
induced network sampling scheme. Given subnetwork size $m$ and number
of bootstrap iterates $B$, the \textit{uniform subsampling} bootstrap
scheme has the following steps:
\begin{longlist}[(1)]
\item[(1)] For the $b$th iterate of the bootstrap, $b = 1, \ldots, B$.
\item[(2)] Choose $m$ vertices without replacement from $V(G)$ and form
the induced subgraph of $G$ based on the selected vertices. Denote the
graph formed by $H$.
\item[(3)] Calculate $\hat{P}_{b1}(R)$, given by formula
%
\begin{equation}
\label{uniform} \hat{P}_{b1}(R) = \frac{1}{{m \choose p} |\operatorname{Iso}(R)|}\sum
_{S\subseteq K_m,
S\cong R}\mathbf{1}(S \subseteq H).
\end{equation}
\end{longlist}
The \emph{uniform subsampling} bootstrap estimate of $\hat{P}(R)$ is
given by
%
\begin{equation}
\label{eq_unif} \bar{P}_{B1}(R) = \frac{1}{B}\sum
_{b = 1}^B \hat{P}_{b1}(R).
\end{equation}

The uniform subsampling bootstrap scheme is the network version of the
common subsampling bootstrap scheme seen in Bickel et al. \cite
{MR1441142}. Note that there are other ways of forming uniformly
subsampled bootstrap estimates, as mentioned in~\cite{MR1441142};
however, we just mention one in this discourse.

For the bootstrap method, we prove a theorem of following type:
%
\begin{thm}
\label{int_thm1}
Suppose $R$ is fixed acyclic or $p$-cycle with $|V(R)| = p$ and $|E(R)|
= e$:
\begin{longlist}[(iii)]
\item[(i)] given $G$, $\hat{P}_{b1}(R)$ is an unbiased estimate of
$\hat{P}(R)$;
\item[(ii)] given $G$, $\var (\rho_n^{-e}\hat
{P}_{b1}(R)|G ) =
O (\frac{1}{m^p\rho_n^e}\vee\frac{1}{m} )$;
\item[(iii)] also, if $B\rar\infty$, $n \rar\infty$, $m \rar
\infty$,
$m/n \rar0$ and $B(m^p\rho_n^e\wedge m) > O(n)$, under $G$ generated
from \eqref{eq:nonpar_model},
%
\begin{equation}
\sqrt{n} \bigl(\rho_n^{-e}\bar{P}_{B1}(R) -
\rho_n P(R) \bigr) \stackrel {P} {\rar} 0.
\end{equation}
\end{longlist}
%
\end{thm}

\begin{pf}
The proof is given in Appendix A2 in \cite{bhattacharyya2015supplement}.
\end{pf}

\subsection{Subgraph subsampling bootstrap}
\label{sampling}
In the \emph{subgraph subsampling} bootstrap scheme, we use an
enumeration scheme to find all possible subgraphs $R$ of size $|V(R)| =
p$ in the graph $G$. Then we convert the enumeration scheme into a
sampling scheme by selecting each subgraph $R$ of size $p$ of $G$ with
a fixed probability and counting the number of sampled subgraphs. The
enumeration scheme was proposed by Wernicke et al. \cite
{wernicke2006efficient}. A random version of the enumeration scheme was
also proposed in \cite{wernicke2006efficient}. We use the random
version of the enumeration scheme to form our sampling scheme.

\begin{algorithm}[b]
\caption{\textsc{AssignOrder}($G, p$)}
\label{ao}
\begin{algorithmic}[1]
\Require A graph $G = (V, E)$, where $|V(G)| = n$.
\Ensure A vector $\gs=(\gs(1), \ldots, \gs(n))$, where $\gs$ is some
permutation of $\{1, \ldots, n\}$, and $\gs(i)$ is associated with
vertex $v_{\gs(i)}\in V(G)$ for all $i = 1, \ldots, n$.
\State$\gs_1 \leftarrow1$
\State$\mathcal{V} \leftarrow\{v_1\}$
\State$i \leftarrow1$
\While{$|\mathcal{V}| < n$}
\State Denote $k \leftarrow|N(\mathcal{V})\setminus\mathcal\cV|$
and $\{v_{h_1}, \ldots, v_{h_k}\} = N(\mathcal{V})\setminus\cV$
\State Define $\gs(i + j) \leftarrow h_j$ for $j = 1, \ldots, k$.
\State$i \leftarrow i+k$
\State$\mathcal{V} \leftarrow\mathcal{V}\cup N(\mathcal{V})$
\EndWhile
\end{algorithmic}
\end{algorithm}

Let us first discuss the enumeration scheme of Wernicke et al. \cite
{wernicke2006efficient}, which we shall henceforth call
\texttt{ESU}. The enumeration algorithm is a \emph{breadth-first
search} algorithm. The enumeration scheme creates a forest of tree
structures such that each tree corresponds to one vertex of the network
$G$, and each leaf of each tree is a size-$p$ subgraph [we have $|V(R)|
= p$] of $G$. Since the counting scheme follows a breadth-first search
route, before performing the \texttt{ESU} algorithm, we need an
ordering of the vertices based on breadth-first search of the graph
starting from any particular vertex (say, $v_1$). We get such a
particular fixed ordering of the vertices of the network with $v_1$
getting lowest order value and subsequently, searched vertices getting
higher order values. The ordering is described in the algorithm
\textsc{Assign Order} or \texttt{AO} \ref{ao}, where, given any set of vertices $\cV$, we
denote the set of vertices connected to $\cV$, that is, the \emph
{neighbors} of $\cV$, by $N(\cV)$. Also, based on the ordering defined
by \texttt{AO}, we denote $v_i \succ v_j$, if $v_i$ has a higher order
than $v_j$.

The enumeration algorithm
starts with an available vertex of lowest possible order (where order
is specified by Algorithm \texttt{AO} \ref{ao}), say $v_1$. We construct\vadjust{\goodbreak} a tree
with the vertex $v_1$ as the root node. We consider $v_1$ as the
``parent'' node and neighbors of $v_1$, which have higher order than
$v_1$, as its ``children.'' In the next step, the ``children'' node
becomes the ``parent'' node in the tree and has its own neighbors, which
have higher order than the nodes that have already come into the tree
as their ``children.'' We define $N_{\mathrm{excl}}(v, \cV)$ ($v$ is a vertex,
and $\cV$ is a set of vertices) for $N(v)\setminus\cV$. The tree is
allowed to grow up to a height $p$ if we are counting size-$p$
subgraphs. Thus we can see that each leaf of the tree represents a
collection of $p$ nodes coming from the path connecting the leaf to the
root $v_1$. For each vertex, we have such a tree, and over counting is
averted as we maintain the order of vertices assigned by Algorithm \texttt{AO}
\ref{ao} while forming the trees. So, with the help of the particular
ordering of vertices, each of the size-$p$ subgraphs ($|V(R)| = p$) is
counted only once.

The randomized enumeration Algorithm \texttt{RAND-ESU} \ref{rand-esu}
also creates a forest of tree structures such that each tree
corresponds to one vertex of the network $G$, and each leaf of each
tree is a size-$p$ subgraph [we have $|V(R)| = p$] of $G$. However,
only a random selection of leaves of each tree is present in \texttt
{RAND-ESU} with uniform probability of selection of each leaf. The
random enumeration algorithm
starts with an available vertex of lowest possible order (where, order
is specified by Algorithm \texttt{AO} \ref{ao}), say $v_1$, chosen with
probability $q_1$. We construct a tree with the vertex $v_1$ as the
root node. We consider $v_1$ as the ``parent'' node and neighbors of
$v_1$, which have a higher order than $v_1$ as its ``children'' and each
``child'' is selected with probability $q_2$ independently. In the next
step, the ``children'' nodes become the ``parent'' nodes in the tree and
has their own neighbors, which have higher order than the nodes that
have already come into the tree, as their ``children,'' and each
``child'' is selected with probability $q_3$. The tree is allowed to
grow up to a height $p$ if we are counting size-$p$ subgraphs, and at
step $d$, the probability of selection is $q_d$. So we can see that
each leaf of the tree represents a collection of $p$ nodes coming from
the path connecting the leaf to the root. For each vertex, we have such
a tree. So, with the help of the particular ordering of vertices, a
subsample of the size-$p$ subgraphs ($|V(R)| = p$) is obtained. The
pseudo-code is given in Algorithm \ref{rand-esu}.

The ordering is needed for success of the \texttt{ESU} algorithm and
its randomized counterpart \ref{rand-esu}. We formally state the
subsampling algorithm, \texttt{RAND-ESU} \ref{rand-esu} in this paper
with an extra set of parameters $(q_1, \ldots, q_p)$. The enumeration
version can be found in \cite{wernicke2006efficient}.

%
\begin{algorithm}[t]
\caption{\textsc{RandomizedEnumerateSubgraph}($G, p$)}
\label{rand-esu}
\begin{algorithmic}[1]
\Require A graph $G = (V, E)$, an integer $p$ and an vector $(q_1,
\ldots, q_p)$, where $1\leq p \leq|V|$ and $q_d \leq1$ for all $d =
1, \ldots, p$.
\Ensure$\s^R_p = $ A sample of subgraphs, $R$ of $G$, such that $|R|
= p$.
\For{each vertex $v\in V$}
\State$V_{\mathrm{Extension}} \leftarrow\{u\in N(\{v\}): u\succ v\}$
\State$d \leftarrow1$
\State\textbf{With probability $q_d$} \textbf{Call}
RandExtendSubgraph($\{
v\}, V_{\mathrm{Extension}}, v, d\}$)
\EndFor
\Function{ \textsc{RandExtendSubgraph}}{$V_{\mathrm{Subgraph}}, V_{\mathrm{Extension}},
v, d$}
\State\textbf{Input:} Graphs $V_{\mathrm{Subgraph}}, V_{\mathrm{Extension}}$ and vertex $v$.
\State\textbf{Output:} A sample of subgraphs, $R$ of $G$, such that
$|V(R)| = p$ and $v$ is a vertex of $R$.
\If{$|V_{\mathrm{Subgraph}}| = p$}
\State\Return Subgraph of $G$ induced by $V_{\mathrm{Subgraph}}$
\Else
\While{$V_{\mathrm{Extension}} \neq\phi$}
\State Remove an arbitrarily chosen vertex $w$ from $V_{\mathrm{Extension}}$
\State$V'_{\mathrm{Extension}} \leftarrow V_{\mathrm{Extension}} \cup\{u\in N_{\mathrm{excl}}(w,
V_{\mathrm{Subgraph}}): u\succ v\}$
\State$d \leftarrow|V_{\mathrm{Subgraph}}| + 1$
\State\textbf{With probability $q_d$} \textbf{Call}
RandExtendSubgraph($V_{\mathrm{Subgraph}}\cup\{w\}, V'_{\mathrm{Extension}}, v, d$)
\EndWhile
\EndIf
\State\Return
\EndFunction
\end{algorithmic}
\end{algorithm}
%

From the sampling scheme \texttt{RAND-ESU} we have a sample $\s^R_p$ of
size-$p$ subgraphs of $G$. Now, if we consider each item to be one
size-$p$ subgraph of $G$, that is, an element of $\s_p$, then we can
try to calculate the \emph{inclusion probability} of each item in the
sample $\s^R_p$.\vadjust{\goodbreak}

The item $S\in\s_p$ is a subgraph of $G$ induced by the set of
vertices $\{w_1, \ldots, w_p\}$, where we take that $w_{i + 1} \succ
w_i$, $i = 1, \ldots, p-1$. Thus:
\begin{eqnarray*}
\pi\equiv\mbox{Inclusion probability of } S & = & \Pr\bigl[(w_1,
\ldots, w_p) \mbox{ is selected}\bigr]
\\
& = & \Pr\bigl[w_p|(w_1, \ldots, w_{p - 1})
\mbox{ is selected}\bigr]
\\
&&{}\times\Pr\bigl[(w_1, \ldots, w_{p - 1}) \mbox{ is selected}
\bigr]
\\
& = & q_p\cdot\Pr\bigl[(w_1, \ldots, w_{p - 1})
\mbox{ is selected}\bigr]
\\
& = & q_p\cdot q_{p - 1}\cdot\Pr\bigl[(w_1,
\ldots, w_{p - 2}) \mbox{ is selected}\bigr]
\\
& = & \cdots= q_p\cdot q_{p - 1}\cdots q_1 =
\prod_{d = 1}^p q_d.
\end{eqnarray*}

So, each item $S\in\s_p$ has an inclusion probability $\pi$ to be in
the sample $\s^R_p$.

In Theorem~2 of \cite{wernicke2006efficient} it was proved that the
output of the \texttt{ESU} algorithm $\s_p$ contains all subgraphs $R$
of $G$, such that $|V(R)| = p$, exactly once. Thus we can write statistic
(\ref{statistic}) for a specific subgraph $R$ with $|V(R)| = p$ in the
following way:
%
\begin{equation}
\label{statistic1} \hat{P}(R) = \frac{1}{{n \choose p} }\sum_{S\in\s_p}
\mathbf{1}(S \cong R).
\end{equation}
Essentially, we have a normalized \emph{population total} in terms of
sampling theory. Our goal is to form a sampling design and devise a
corresponding sampling estimator of $\hat{P}(R)$ given $G$. To meet
this goal we use a sampling version of the enumeration scheme \texttt{ESU}.

Now we have a sampling scheme by which we select a sample $\s^R_p$ from
the population $\s_p$, where each element of $\s_p$ has probability of
inclusion of $\pi$. Thus we can define a \emph{Horvitz--Thompson}
estimator (for reference, see Chapter~6.2 of~\cite{MR2894042}) of
$\hat
{P}(R)$ based on $\s^R_p$ as
%
\begin{equation}
\label{eq_sampling} \hat{P}_{b2}(R) = \frac{1}{ (\prod_{d=1}^p q_d ){n
\choose p}
}\sum
_{S\in\s^R_p}\mathbf{1}(S \cong R).
\end{equation}

Now if we repeat the same procedure $B$ number of times, each time
getting independent copies of $\s^R_p$ with replacement from $\s_p$, we
can get the \emph{subgraph subsampling} bootstrap estimate,
%
\begin{equation}
\label{eq_sampling2} \bar{P}_{B2}(R) = \frac{1}{B} \sum
_{b=1}^B \hat{P}_{b2}(R).
\end{equation}


For the bootstrap method, we prove a theorem of following type:
%
\begin{thm}
\label{int_thm2}
Suppose $R$ is fixed acyclic or $p$-cycle with $|V(R)| = p$ and $|E(R)|
= e$:
\begin{longlist}[(iii)]
\item[(i)] given $G$, $\hat{P}_{b2}(R)$ is an unbiased estimate of
$\hat{P}(R)$;
\item[(ii)] given $G$, $\var (\rho_n^{-e}\hat
{P}_{b2}(R)|G ) =
O ( (\frac{1}{q_1}-1 )\frac{1}{n} + \frac{1}{n\rho
_n^{e-p+1}}\cdot\prod_{d=2}^p\frac{1}{\gl_nq_d} )$;
\item[(iii)] for $B \rar\infty$ and $q_d \rar0$ for all $d=1,\ldots
,p$ such that $\frac{1}{B} (\frac{1}{q_1} - 1 ) \rar0$
and $B\prod_{d=2}^pq_d \geq\frac{1}{n^{p-1}\rho_n^e}$ and $n \rar
\infty$, $\gl_n \rar\infty$, and under $G$ generated from \eqref
{eq:nonpar_model},
%
\begin{equation}
\label{eq_sample_boot_thm} \sqrt{n}\bigl(\rho_n^{-e}
\bar{P}_{B2}(R) - \rho_n^{-e}P(R)\bigr)\stackrel
{P} {\rar } 0.
\end{equation}
\end{longlist}
%
\end{thm}

\begin{pf}
The proof is given in Appendix A3 in \cite{bhattacharyya2015supplement}.
\end{pf}

Note that the main reason for taking repeated independent samples, $\s
_p^R$ for this case, is to reduce the variance of the bootstrap
estimates and to make the estimates more stable.

\subsection{Estimation of variance and covariance}
\label{sec_est_var}

We first start with the situation when the source of variation is only
the randomness coming from sampling from the underlying model \eqref
{eq:nonpar_model}. We denote $\var [\rho^{-e}\hat{P}(R)
]$ as
$\gs^2(R;\rho)$ and $\cov (\rho^{-e_1}\hat{P}(R_1), \rho
^{-e_2}\hat
{P}(R_2) )$ as $\gs(R_1, R_2;\rho)$. Note that, $e_1=|E(R_1)|$,
$e_2 = |E(R_2)|$, $p_1 = |V(R_1)|$ and $p_2 = |V(R_2)|$.

\begin{proposition}
\label{prop_var_est}
For connected subgraphs $R$, $R_1$ and $R_2$ of $G$, we have that
\begin{eqnarray*}
\gs^2(R;\rho) & = & \frac{1}{ (\rho^{e}{n \choose p}
|\operatorname{Iso}(R)|
)^2}\mathop{\sum
_{W: W =S\cup T}}_{ S, T\cong R, S \cap T \neq
\varnothing}\E \biggl[\sum
_{W\subseteq K_n}\mathbf{1}(W \subseteq G) \biggr]
\\
& & {}- \biggl(1 - \frac{ ((n-p)! )^2}{n!(n-2p)!} \biggr) \bigl(\tilde{P}(R)
\bigr)^2,
\\
\gs(R_1, R_2;\rho) &=& \frac{1}{ (\rho^{e_{1}+e_{2}}{n \choose p_1}{n
\choose p_2} |\operatorname{Iso}(R_1)||\operatorname{Iso}(R_2)| )}
\\
&&{}\times\mathop{\sum_{W: W=S\cup T, }}_{ S\cong R_1, T\cong R_2, S\cap T \neq
\varnothing}\E \biggl[\sum
_{W\subseteq K_n}\mathbf{1}(W \subseteq G) \biggr]\\
&&{} - \biggl(1 -
\frac{(n-p_1)!(n-p_2)!}{n!(n-p_1-p_2)!} \biggr)\tilde {P}(R_1)\tilde{P}(R_2).
\end{eqnarray*}
\end{proposition}

\begin{pf}
The proof is given in Appendix B1 in \cite{bhattacharyya2015supplement}.
\end{pf}

Note that if we take $k = |V(W)|$ and $e_W \equiv|E(W)|$, then $k =
p,\ldots, 2p - 1$ and each term of sum on the RHS of the previous
equation is
%
\begin{eqnarray}
\label{var_approx}
 \frac{1}{ (\rho^{e}{n \choose p} |\operatorname{Iso}(R)| )^2}\E
 \biggl[\sum_{W\subseteq K_n}
\mathbf{1}(W \subseteq H) \biggr] &=& \frac{\rho^{e_W}{n \choose k} |\operatorname{Iso}(W)|}{ (\rho^{e}{n \choose p}
|\operatorname{Iso}(R)| )^2} \tilde{P}(W)
\nonumber
\\[-8pt]
\\[-8pt]
\nonumber
&=& O\bigl(n^{k - 2p}\rho^{e_W - 2e}\bigr).
\end{eqnarray}
We can analyze each such term separately:
\begin{longlist}[(1)]
\item[(1)] If $k = |V(W)| = 2p - 1$, then $W$ is a connected graph,
with $e_W = 2e$. Thus we have that the main leading term equals
$O(\frac
{1}{n})$.
%
\item[(2)] In the case $k=|V(W)| < (2p-1)$:
\begin{itemize}
\item If $R$ is acyclic, then $e_W-2e\leq k-2p-1$ since $e=p-1$, so
$O(n^{k - 2p}\times \rho^{e_W - 2e}) = o (n^{-1} )$ if $\gl_n =
n\rho
_n \rar\infty$.
\item If $R$ is a $p$-cycle, $e_W-2e = k-2p < 0$ if $k=|V(W)|=p$ and
$e_W-2e\leq k-2p-1$ if $p < k < (2p-1)$, so $O(n^{k - 2p}\rho^{e_W -
2e}) = O (\gl_n^{-p} ) + o (n^{-1} )$ if $\gl_n
\rar
\infty$.
\item If $R$ is any other cyclic graph, $O(n^{k - 2p}\rho^{e_W - 2e}) =
O (n^{-c}\rho^{-d} )$, where $0<c\leq p$ and $0<d \leq
c(c-1)/2$ for each $c$. So, in order to have $n^{-c}\rho^{-d} \leq
Mn^{-1}$, the worst rate that $\gl_n$ can have is $\gl_n = O
(n^{1-2/p} )$.
\end{itemize}
\end{longlist}
%

For connected and acyclic or $p$-cycle $R$, $R_1$ and $R_2$, we get that
\begin{eqnarray*}
\gs^2(R;\rho) &=& O \biggl(\frac{1}{n}\vee\frac{1}{\gl_n^p}
\biggr),
\\
\gs(R_1, R_2;\rho) &=& O \biggl(\frac{1}{n} \biggr).
\end{eqnarray*}

So, for calculation of variance, if $R$ is acyclic or $p$-cycle, we
only estimate the count of the features which are $W = S\cup T$ and
$|V(W)| = 2p - 1$ and $|V(W)| = p$.
Thus using the expansion given in Proposition~\ref{prop_var_est}, the
empirical estimator of $\gs^2(R;\rho)$ is defined as
%
\begin{equation}\qquad
\label{eq_var_est} \hat{\gs}^2(R) = \frac{1/(1-x)}{ (\hat{\rho}_n^{e}{n \choose p}
|\operatorname{Iso}(R)| )^2}\mathop{\sum
_{W\subseteq K_n: W=S\cup T,}}_{ S,
T\cong R, |S\cap T| =1,p}\mathbf{1}(W \subseteq G) -
\frac{x\hat
{\rho
}_n^{-2e}\hat{P}(R)^2}{(1-x)},
\end{equation}
%
where $x =  (1 - \frac{ ((n-p)! )^2}{n!(n-2p)!}
)$, and
%
using the expansion given in Proposition~\ref{prop_var_est}, the
empirical estimator of $\gs(R_1, R_2;\rho)$ is defined as
%
\begin{eqnarray}
\label{eq_cov_est} \hat{\gs}(R_1, R_2) &=& \frac{1/(1-y)}{ (\hat{\rho}_n^{e_{1}+e_{2}}{n
\choose p_1}{n \choose p_2} |\operatorname{Iso}(R_1)||\operatorname{Iso}(R_2)| )}
\nonumber
\\[-8pt]
\\[-8pt]
\nonumber
&&{}\times\mathop{\sum_{W\subseteq K_n, W=S\cup T, }}_{ S\cong R_1, T\cong R_2,
|S\cap T| =1}\mathbf{1}(W
\subseteq G) - \frac{y\hat{\rho}_n^{-(e_1+e_2)}\hat{P}(R_1)\hat{P}(R_2)}{(1-y)},
\end{eqnarray}
where $y =  (1 - \frac{(n-p_1)!(n-p_2)!}{n!(n-p_1-p_2)!} )$.
%

$\hat{\gs}^2(R)$ and $\hat{\gs}(R_1,R_2)$ become consistent
estimates of
$\gs^2(R;\rho)$ and $\gs(R_1, R_2; \rho)$ as well as $\gs^2(R;\hat
{\rho
})$ and $\gs(R_1, R_2;\hat{\rho})$, respectively.

\begin{lemma}
\label{lem_var_est1}
As $\gl_n \rar\infty$ and $n\rar\infty$, if $R$, $R_1$, $R_2$ is
connected acyclic or  $p$-cycle, then additionally $\gl_n^p \geq O(n)$,
%
\begin{eqnarray}
\frac{\hat{\gs}^2(R)}{\gs^2(R;\hat{\rho})} &\stackrel{P} {\rar}& 1,
\\
\frac{\hat{\gs}(R_1, R_2)}{\gs(R_1, R_2; \hat{\rho})} &\stackrel {P} {\rar }& 1.
\end{eqnarray}
\end{lemma}

\begin{pf}
The proof is given in Appendix B2 in \cite{bhattacharyya2015supplement}.
\end{pf}

Now we can see that $\hat{\gs}^2(R)$ and $\hat{\gs}(R_1,R_2)$ are
nothing but count statistics on the statistic $W=S\cup T$. So, using
bootstrap methods, we define a bootstrap-based estimate of $\hat{\gs
}^2(R)$, for $i=1,2$,
%
\begin{eqnarray}
\label{eq_boot_var} \hat{\gs}_{Bi}^2(R)& =& \mathop{\sum
_{W=S\cup T, S, T\cong R,}}_{
|S\cap
T| =1,p}\frac{ (\hat{\rho}_n^{e_W}{n \choose p_W}
|\operatorname{Iso}(R)|
)}{(1-x) (\hat{\rho}_n^{e}{n \choose p} |\operatorname{Iso}(R)| )^2}\bar
{P}_{Bi}(W)
\nonumber
\\[-8pt]
\\[-8pt]
\nonumber
&&{}- \frac{x\hat{\rho}_n^{-2e}\bar{P}_{Bi}(R)^2}{(1-x)},
\end{eqnarray}
where $x =  (1 - \frac{ ((n-p)! )^2}{n!(n-2p)!}
)$. A
bootstrap-based estimate of $\hat{\gs}(R_1, R_2)$ is
%
\begin{eqnarray}\qquad
\label{eq_boot_cov} &&\hat{\gs}_{Bi}(R_1, R_2) \nonumber\\
&&\qquad=
\mathop{\sum_{W=S\cup T, S\cong R_1,}}_{
T\cong R_2, |S\cap T| =1}
\frac{ (\hat{\rho}_n^{e_W}{n \choose
p_W} |\operatorname{Iso}(W)| )}{(1-y) (\hat{\rho}_n^{e_{1}+e_{2}}{n
\choose
p_1}{n \choose p_2} |\operatorname{Iso}(R_1)||\operatorname{Iso}(R_2)| )}\bar{P}_{Bi}(W)
\\
&&\qquad\quad{}- \frac{y\hat{\rho}_n^{-(e_1+e_2)}\bar{P}_{Bi}(R_1)\bar
{P}_{Bi}(R_2)}{(1-y)},\nonumber
\end{eqnarray}
where $y =  (1 - \frac{(n-p_1)!(n-p_2)!}{n!(n-p_1-p_2)!} )$
and $\bar{P}_{Bi}(W)$ ($i=1,2$) are bootstrap count statistics
estimates, defined in equations \eqref{eq_unif} and \eqref{eq_sampling2}.

\begin{lemma}
\label{lem_boot_var}
As $\gl_n \rar\infty$, $n\rar\infty$, $B \rar\infty$ and under the
conditions of Theorems \ref{int_thm1} and~\ref{int_thm2}, if
$R$, $R_1$ and $R_2$ are acyclic or $p$-cycle, then additionally $\gl_n^p
\geq O(n)$,
%
\begin{eqnarray}
\frac{\hat{\gs}_{Bi}^2(R)}{\gs^2(R;\hat{\rho})} &\stackrel {P} {\rar}
 &1\qquad \mbox{for } i=1,2,
\\
\frac{\hat{\gs}_{Bi}(R_1, R_2)}{\gs(R_1, R_2;\hat{\rho})} &\stackrel {P} {\rar} &1 \qquad\mbox{for } i=1,2.
\end{eqnarray}
\end{lemma}

\begin{pf}
The proof is given in Appendix B3 in \cite{bhattacharyya2015supplement}.
\end{pf}

\subsection{Comparison of the bootstrap methods}
The variance of each of the subsampling bootstrap methods, just on the
basis of the randomness generated from the bootstrap sampling, is given
in Theorems \ref{int_thm1} and \ref{int_thm2}. Also, the worst-case
computational complexity of finding count statistics for subgraphs $R$
of size $p$, for the \emph{uniform subsampling} bootstrap, becomes
$O (Bm^p )$, whereas for the \emph{subgraph subsampling}
bootstrap scheme, the worst-case complexity is $O (B\prod_{d=1}^p(nq_d) )$. Now the question of balancing computational
complexity and statistical stability become important.

For dense networks, say when $\rho_n = n^{-\gve}$ with $\gve>0$ small
(say between $0<\gve<1/2$), we also have $\gl_n = n^{1-\gve}$:
\begin{itemize}
\item For \emph{uniform subsampling} from Theorem \ref{int_thm1}, we get
that $\var (\rho^{-e}\bar{P}_{B1}(R) ) = O (\frac
{1}{n^{1+2\gve}} )$ with $m=n^{\gve}$ and $B=n^{1+\gve}$. The
worst-case computational cost becomes $O (n^{1+(p+1)\gve} )$.
\item For \emph{subgraph subsampling} from Theorem \ref{int_thm2}, we
get that $\var (\rho^{-e}\bar{P}_{B2}(R) ) = O
(\frac
{1}{n^{1+\gve}} )$ for $p$-cycle $R$ and $O (\frac
{1}{n^{1+2\gve}} )$ for acyclic $R$ with $q_d=O (\frac
{1}{n^{1-\gve}} )$ for $d=2,\ldots,p$ and $B=n^{2\gve}$. The
worst-case computational complexity becomes $O (n^{1+(p+1)\gve
} )$.
\end{itemize}
Thus in \emph{dense} networks, both the subsampling bootstrap methods
can achieve low enough bootstrap variance for low computational cost.
In fact, the gain in computational complexity is quite astonishing as
polynomial complexity gets reduced to near-linear complexity. The \emph
{uniform subsampling} bootstrap is a better choice for its ease of use
and marginally smaller variance for $p$-cycle $R$. However, since $m$
has to be greater than $p$, for large $R$, the benefit of using the
\emph{uniform subsampling} bootstrap starts to reduce, and in these
cases, the \emph{subgraph subsampling} bootstrap might be a better choice.

For the sparse case, say when $\rho_n = n^{\gve-1}$ with $\gve>0$ small
(say between $0<\gve<1/2$), we also have $\gl_n = n^{-\gve}$:
\begin{itemize}
\item For \emph{uniform subsampling} from Theorem \ref{int_thm1}, we get
that $\var (\rho^{-e}\bar{P}_{B1}(R) ) = O (\frac
{1}{n^2} )$ for acyclic $R$ and $O (\frac{1}{n^{1+\gve
}}
)$ for $p$-cycle $R$ with $m=n^{1-\gve}$ and $B=n^{1+\gve}$. The
worst-case computational cost becomes $O (n^{p-((p-1)\gve-
1)} )$.
\item For \emph{subgraph subsampling} from Theorem \ref{int_thm2}, we
get that $\var (\rho^{-e}\bar{P}_{B2}(R) ) = O
(\frac
{1}{n^{2-\gve}} )$ for acyclic $R$ and $O (\frac
{1}{n^{1+\gve
}} )$ for $p$-cycle $R$ with $q_d=O (\frac{1}{n^{\gve
}}
)$ for $d=1,\ldots,p$ and $B=n$. The worst-case computational
complexity becomes $O (n^{p- (p\gve- 1)} )$.
\end{itemize}
Thus in \emph{sparse} networks, the computational advantage of using
the subsampling bootstrap starts to reduce, especially for small
subgraphs $R$. However, for large subgraphs $R$, there is still a
computational advantage to using subsampling bootstrap methods. The
\emph{subgraph subsampling} bootstrap scheme is a better choice in this
case as it has smaller variance for similar computational complexity.

But for \emph{sparse} graphs, the methods still remain polynomial in
worst-case complexity, and for large $p$ and $n$, the methods become
numerically infeasible. In those cases, it becomes more of a \emph
{detection} problem than a \emph{counting} problem, and a fundamentally
different approach will be required for feasible inference.

%
%


\section{Theoretical results}
\label{sec_theory}
In this section, we shall try to provide asymptotic distribution for
normalized bootstrap estimates of count statistics. We define
normalized bootstrap estimates of count statistic for subgraph $R$ from
\eqref{eq_unif} and \eqref{eq_sampling2} by
%
\begin{equation}
\label{eq_boot_est} \hat{T}_{Bi}(R) = \hat{\rho}_n^{-e}
\bar{P}_{Bi}(R),
\end{equation}
where $i=1,2$ for the two different bootstrap schemes.
By obtaining an estimate of the asymptotic variance of $\rho^{-e}\hat
{P}(R)$, we can estimate its asymptotic distribution and thus construct
hypothesis tests based on the asymptotic distribution.
We combine the results obtained in Section~\ref{methods} to prove
Theorem~\ref{gen_thm}.

\begin{thm}
\label{gen_thm}
Suppose $R$ is fixed, acyclic or $p$-cycle with $|V(R)| = p$ and
$|E(R)| = e$ and $\int_0^\infty\int_0^\infty w^{2e}(u, v) \,du\,dv <
\infty$. Under the conditions defined in Theorems \ref{int_thm1} and
\ref{int_thm2}, for $i=1,2$, if $\gl_n(\equiv n\rho_n) \rar\infty$ and
$B\rar\infty$,
%
\begin{eqnarray}
\sqrt{n} \bigl(\hat{T}_{Bi}(R) - \til{P}(R) \bigr) &\stackrel {P} {
\rar} & 0,
\\
\sqrt{n} \biggl(\frac{\hat{T}_{Bi}(R) - \tilde{P}(R)}{\hat{\gs
}_{Bi}(R)} \biggr) &\stackrel{w} {\rar}& N(0, 1).
\end{eqnarray}
If for fixed, acyclic or $p$-cycle subgraphs\vspace*{1pt} $(R_1, \ldots, R_k)$, we
define,
$\mathbf{T}_{Bi}(\mathbf{R}) =  (\hat{T}_{Bi}(R_1), \ldots,
\hat
{T}_{Bi}(R_k) )$ and $\mathbf{P}(\mathbf{R}) =  (\til{P}(R_1),
\ldots, \til{P}(R_k) )$
%
\begin{equation}
\sqrt{n} \bigl( \bigl(\mathbf{T}_{Bi}(\mathbf{R}) - \mathbf {P}(
\mathbf {R}) \bigr)^T\hat{\Sigma}_{Bi}^{-1/2}(
\mathbf{R}) \bigl(\mathbf {T}_{Bi}(\mathbf{R}) - \mathbf{P}(\mathbf{R})
\bigr) \bigr)\stackrel {w} {\rar} N(\mathbf{0}, \mathbf{I}),
\end{equation}
where $[\hat{\Sigma}_{Bi}]_{st} = \hat{\gs}_{Bi}(R_s,R_t)$, $s, t=1,
\ldots, k$ and if $R_s=R_t=R$, $\hat{\gs}_{Bi}(R_s, R_t) = \hat{\gs
}_{Bi}^2(R)$. These results also hold for subgraphs $R$, which are $r$-cycles.
\end{thm}


\subsection{Proof of Theorem \texorpdfstring{\protect\ref{gen_thm}}{3}}
\label{proof_gen_thm}
The proof follows from the lemma and theorems of the previous section.
Since we have $\sqrt{n}$-consistent bootstrap estimators of $\rho
^{-e}\bar{P}_{Bi}(R)$ for $i=1,2$. Now, from the Theorem~1(a) in \cite
{MR2906868}, we know that as $\gl_n \rar\infty$ if $\hat{\rho}_n =
\frac{\bar{D}}{n-1}$, as defined in \eqref{eq_rho_def},
\begin{eqnarray*}
\frac{\hat{\rho}_n}{\rho_n} &\stackrel{P} {\rar}& 1 ,\\
\sqrt{n} \biggl(\frac{\hat{\rho}_n}{\rho_n} - 1
\biggr)& \stackrel {w} {\rar } &N\bigl(0, \gs^2\bigr).
\end{eqnarray*}
Now, we define the bootstrap estimates in \eqref{eq_boot_est}. Thus we
get by applying Slutsky's Theorem that
\[
\sqrt{n} \bigl(\hat{T}_{Bi}(R) - \til{P}(R) \bigr) \stackrel {P} {
\rar} 0 \qquad\mbox{for } i=1,2.
\]
The statement about bootstrap estimate of variance follows from Lemma~\ref{lem_boot_var} and the definitions of bootstrap variance in the
form of equation \eqref{eq_boot_var}.

Thus we have $\sqrt{n}$-consistent bootstrap estimators, $\hat
{T}_{Bi}(R)$ (for $i=1,2$) of $\hat{T}(R)$ and consistent estimators,
$\hat{\gs}_{Bi}^2(R)$ (for $i=1,2$) of $\gs^2(R;\rho)$. Also from
Theorem~1 of \cite{MR2906868}, we have, for subgraphs $R_1, \ldots,
R_k$ of $G_n$,
\[
\sqrt{n} \bigl( \bigl(\hat{T}(R_1), \ldots, \hat{T}(R_k)
\bigr) - \bigl(\til{P}(R_1), \ldots, \til{P}(R_k) \bigr)
\bigr)\stackrel {w} {\rar} N\bigl(\mathbf{0}, \Sigma(\mathbf{R})\bigr).
\]
Thus we can combine the results from Theorems \ref{int_thm1} and \ref
{int_thm2} with the above theorem, using Slutsky and convergence type
theorems, to get the symptomatic normality behavior of $\hat
{T}_{Bi}(R)$. As $n\rar\infty$, $\gl_n \rar\infty$, and under the
conditions of Theorems~\ref{int_thm1} and \ref{int_thm2}, if we define
$\mathbf{T}_{Bi}(\mathbf{R}) =  (\hat{T}_{Bi}(R_1), \ldots,
\hat
{T}_{Bi}(R_k) )$ and $\mathbf{P}(\mathbf{R}) =  (\til{P}(R_1),
\ldots, \til{P}(R_k) )$
\[
\sqrt{n} \bigl( \bigl(\mathbf{T}_{Bi}(\mathbf{R}) - \mathbf {P}(
\mathbf {R}) \bigr)\hat{\Sigma}_{Bi}^{-1/2}(\mathbf{R}) \bigl(
\mathbf {T}_{Bi}(\mathbf{R}) - \mathbf{P}(\mathbf{R}) \bigr) \bigr)
\stackrel {w} {\rar} N(\mathbf{0}, \mathbf{I}) \qquad\mbox{for } i=1,2
\]
where $[\hat{\Sigma}_{Bi}]_{st} = \hat{\gs}_{Bi}(R_s,R_t)$, $s, t=1,
\ldots, k$, and if $R_s=R_t=R$, $\hat{\gs}_{Bi}(R_s, R_t) = \hat
{\gs
}_{Bi}^2(R)$ for $i=1,2$.

\section{Simulation results}
\label{simulation}
We apply the two representative bootstrap subsampling schemes for
simulated datasets to determine their performance. We generate data
from two different simulation models. Both models are special cases of
the nonparametric model described in \cite{bickel2009nonparametric}.
The two models that we consider are the following:
\begin{itemize}
\item the stochastic block model and
\item the preferential attachment model.
\end{itemize}
For each of the models, we try to find the estimate of the count
statistics features and their confidence intervals through bootstrap
subsampling. The features that we consider are generalized $(k,
l)$-wheels, $p$-cycles and a smooth function of count statistics, transitivity.

\subsection{Count statistics}


In these simulations, the main class of acyclic features we consider
are $(k, l)$-wheels. We also consider the count of the cyclic patterns
such as triads or triangles or 3-cycles and tetrads or \textit
{quadrilaterals} or 4-cycles. We also consider a smooth function of
counts of triangle and $(1,2)$-wheel, known as \textit{transitivity},
$\hat{P}_{\mathrm{Tr}}$, defined in Example~\ref{ex_transitivity}.

\subsection{Stochastic block model}
\label{sec_sbm}
Let $w$ correspond to a $K$-block model defined by parameters $\gt=
(\V
{\pi}, \rho_n, S)$, where $\pi_a$ is the probability of a node being
assigned to block $a$ as before, and
\[
\mathbf{F}_{ab} = \Pr(A_{ij} =1|i\in a,j\in
b)=s_nS_{ab},\qquad 1\leq a,b \leq K,
\]
and the probability of node $i$ to be assigned to block $a$ is $\pi_a$
($a=1, \ldots, K$).

We consider a stochastic block model with $K = 2$, $S =
\bigl(
{
0.4 \atop 0.4}\enskip {0.5 \atop 0.7}
 \bigr)
$, $s_n = \frac{5\nu\sqrt{n}}{n}$ and $\bolds{\pi} = (0.5, 0.5)$. Thus
we get $\rho_n = \V{\pi}^T\mathbf{F}\V{\pi}$. First, we keep $n=1000$
fixed and vary $\nu$ such that $\rho_n$ varies from 10 to 100. Second,
we vary $\nu$ fixed at 0.5 and vary $n=500$ to 3000.

\begin{figure}
\centering
\begin{tabular}{@{}cc@{}}

\includegraphics{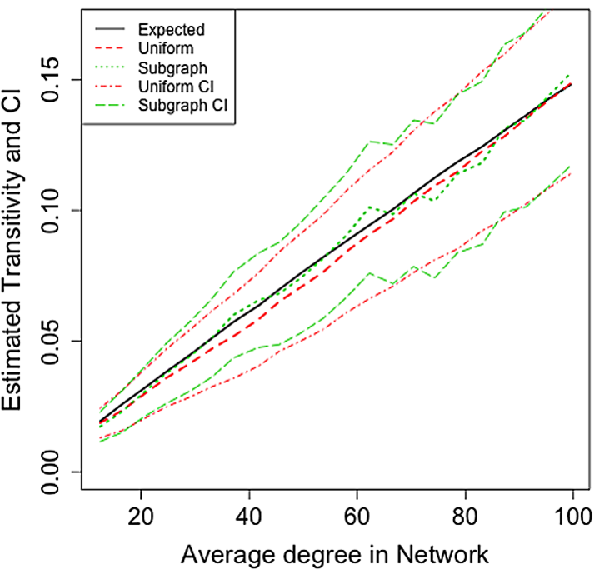}  & \includegraphics{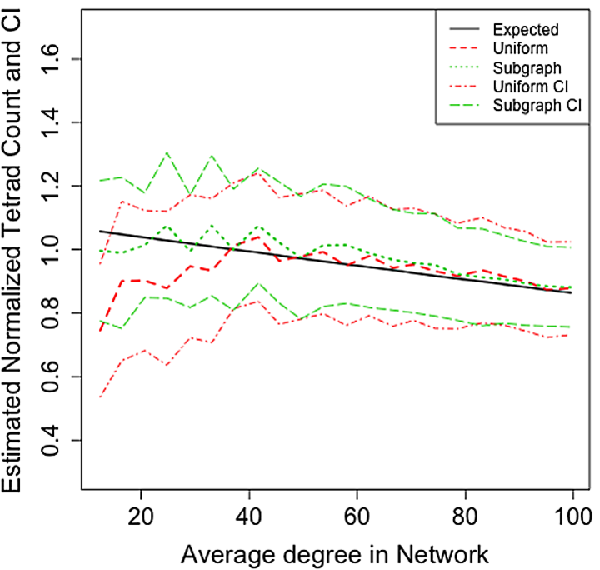}\\
\footnotesize{(a)} & \footnotesize{(b)}\\

\includegraphics{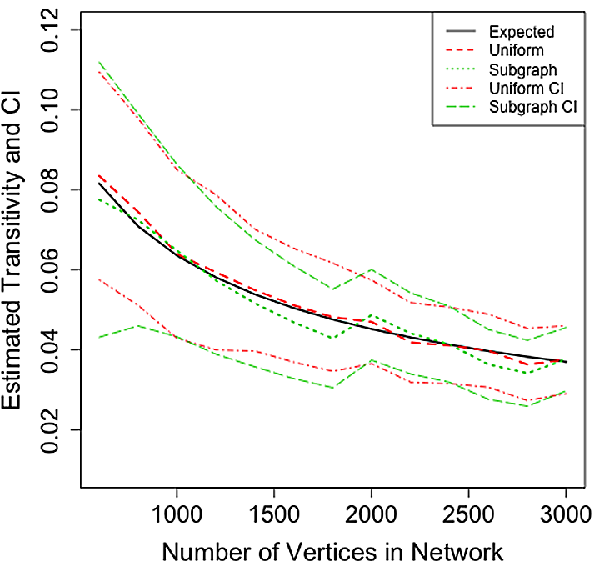}  & \includegraphics{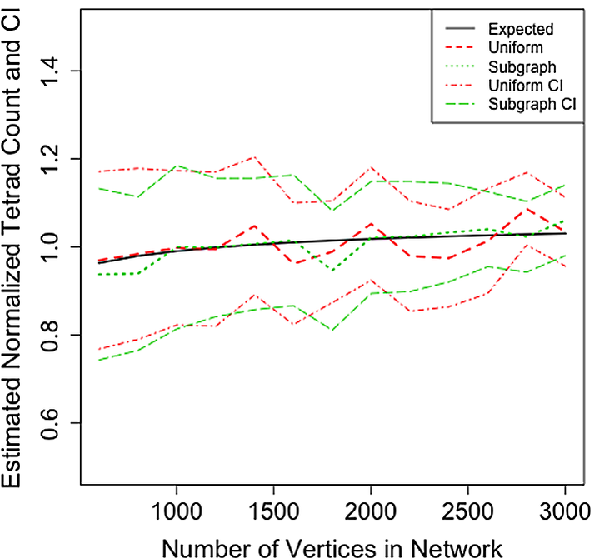}\\
\footnotesize{(c)} & \footnotesize{(d)}
\end{tabular}
\caption{Stochastic block model: For $n = 1000$, we vary average
degree ($\gl_n$) and \textup{(a)} plot estimated normalized tetrad
count and \textup{(b)}
plot estimated transitivity and their 95\% confidence interval (CI),
where CI is estimated using bootstrap estimates of variance of the
estimators. For $\nu= 0.5$, we vary~$n$, and \textup{(c)} plot estimated
normalized tetrad count and \textup{(d)} plot estimated transitivity
and their
95\% confidence interval (CI). We use different colors to indicate
different bootstrap subsampling schemes and graph parameters.}
\label{fig_sbm_12}
\end{figure}

\begin{figure}[t]
\centering
\begin{tabular}{@{}cc@{}}

\includegraphics{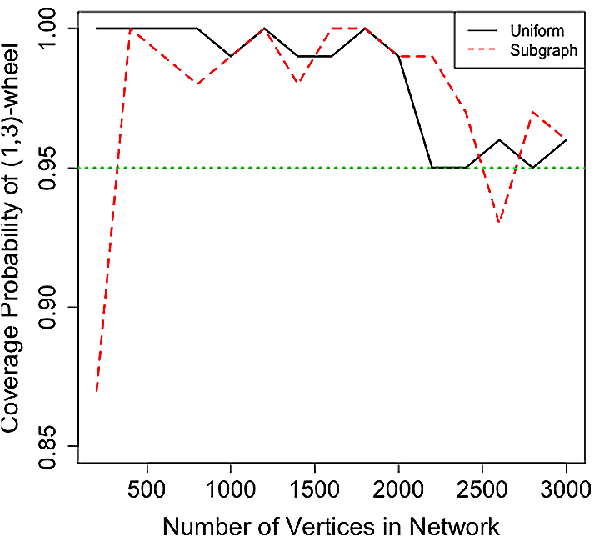}  & \includegraphics{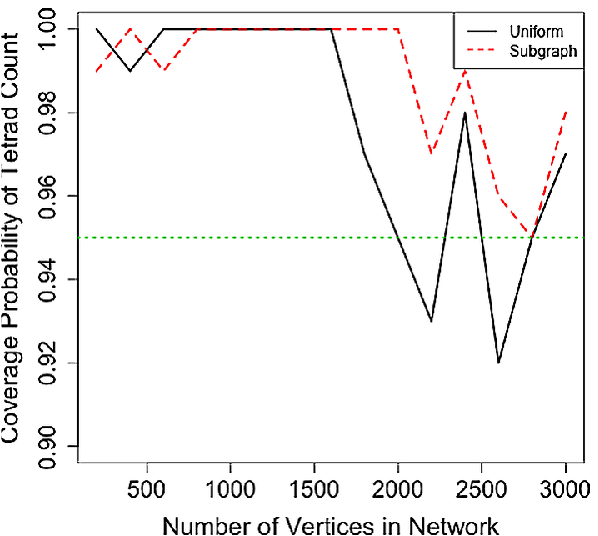}\\
\footnotesize{(a)} & \footnotesize{(b)}
\end{tabular}
\caption{Stochastic block model: For $\nu= 0.5$, we vary $n$
and \textup{(a)} plot estimated coverage probability of 95\% CI for
$(1,3)$-wheel count and \textup{(b)} plot estimated coverage probability of 95\%
CI for normalized tetrad count. We use different colors to indicate
different bootstrap subsampling schemes.}
\label{fig_sbm_cov}
\end{figure}


In the following figures, we try to see the behavior of mean and
variances of the count statistics. In Figure~\ref{fig_sbm_12}(a)--(d),
we compare the asymptotic 95\%\vspace*{1pt} confidence interval of $\til{P}(R)$,
where $R$ is a 4-cycle or tetrad and $\E(\hat{T}_{\mathrm{Tr}})$, using
bootstrap mean and variance estimates, as considered in Theorem~\ref
{gen_thm}. The bootstrap estimate of asymptotic variance of $\hat
{T}^{Bi}_{\mathrm{Tr}}$ is obtained from the bootstrap estimates of $\hat{\gs
}^2_{Bi}(R_1)$, $\hat{\gs}^2_{Bi}(R_2)$ and $\hat{\gs}_{Bi}(R_1,R_2)$
by using Delta method and using the Theorem~\ref{gen_thm}.
%

We also try to see the estimated coverage probabilities of bootstrap
estimated confidence intervals for $\til{P}(R)$. In Figure~\ref
{fig_sbm_cov}(a)--(b), we plot estimated coverage probabilities of
asymptotic 95\% confidence interval for $\til{P}(R)$, where $R$ is a
$(1,3)$-wheel and a 4-cycle. We keep $\nu$ fixed and vary $n$ from 200
to 3000. Estimated coverage probabilities start becoming close to 0.95
at around $n=2000$.

\begin{figure}[t]
\centering
\begin{tabular}{@{}cc@{}}

\includegraphics{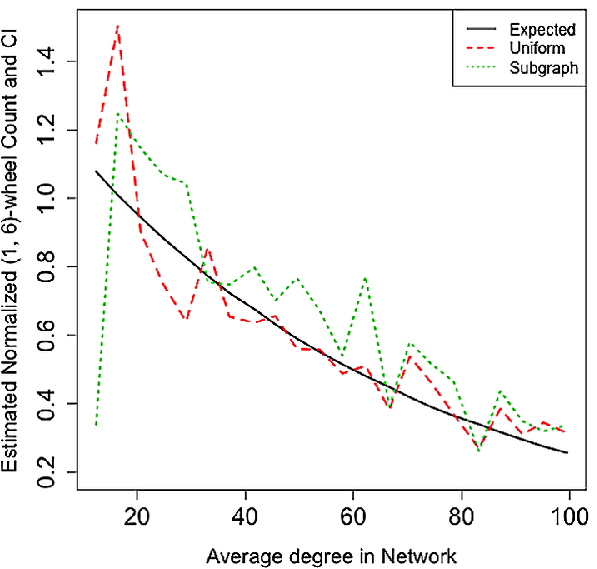}  & \includegraphics{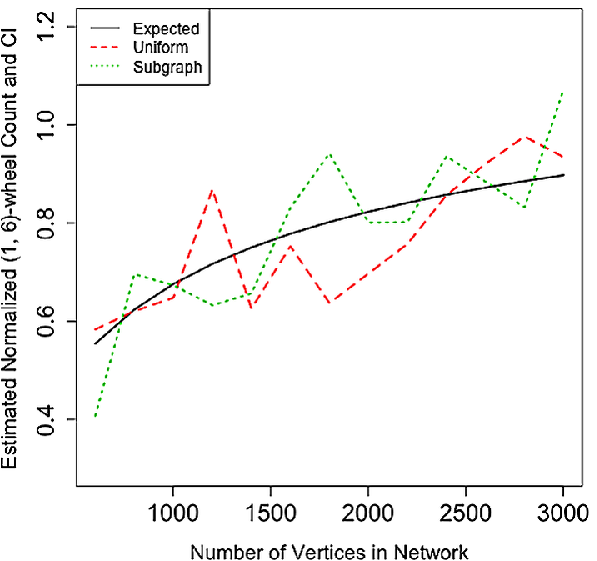}\\
\footnotesize{(a)} & \footnotesize{(b)}
\end{tabular}
\caption{Stochastic block model: For $n = 1000$, we vary average
degree ($\gl_n$) and \textup{(a)} plot estimated normalized $(1,6)$-wheel count.
For $\nu= 0.5$, we vary $n$ and \textup{(b)} plot estimated normalized
$(1,6)$-wheel count. We use different colors to indicate different
bootstrap subsampling schemes and graph parameters.}
\label{fig_sbm_16}
\end{figure}

In Figure~\ref{fig_sbm_16}, we compare the mean of the bootstrap
estimates with the parameter $\tilde{P}(R)$. In Figure~\ref
{fig_sbm_16}(a), we keep $n$ fixed but vary $\gl_n$ from $\gl_n$ 10 to
100 by varying~$\nu$, and in Figure~\ref{fig_sbm_16}(b), we keep $\nu$
fixed and vary $n$ from 500 to 3000. Thus we get reasonable estimates
of integral parameters of graph as we vary the average degree and
number of vertices of the graph.

\begin{figure}[b]
\centering
\begin{tabular}{@{}cc@{}}

\includegraphics{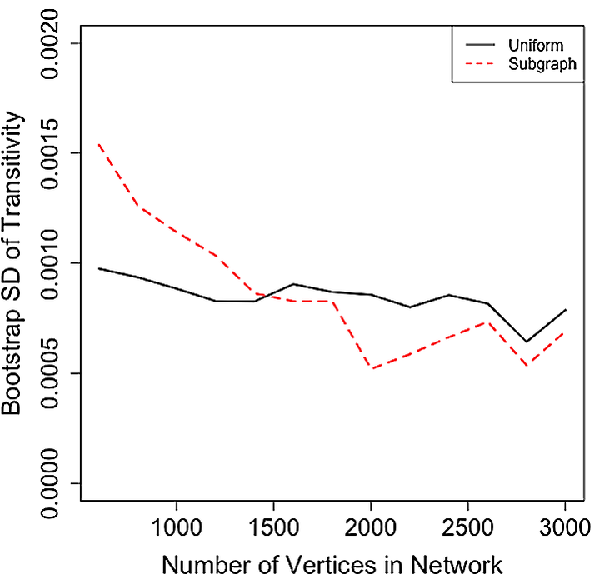}  & \includegraphics{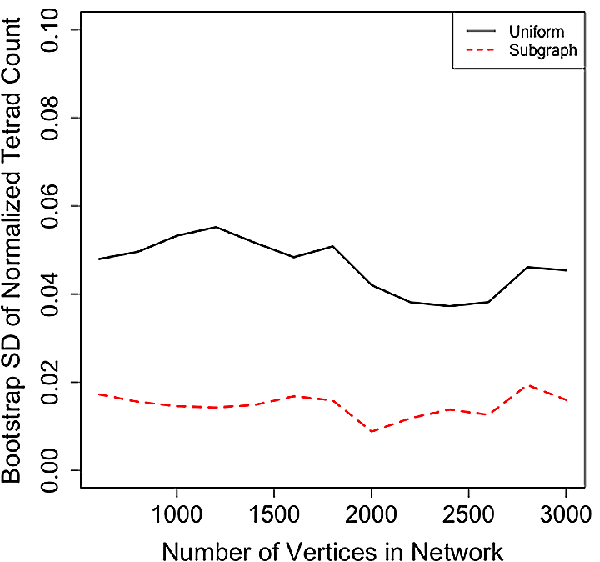}\\
\footnotesize{(a)} & \footnotesize{(b)}
\end{tabular}
\caption{Stochastic block model: For $\nu= 0.5$, we vary the
number of vertices ($n$) and plot \textup{(a)} bootstrap variance of estimated
transitivity and \textup{(b)} bootstrap variance of normalized tetrad
count. We
use different colors to indicate different bootstrap subsampling schemes.}
\label{fig_sbm_var}
\end{figure}

In Figure~\ref{fig_sbm_var}, we compare the variance of the bootstrap
estimates, based on bootstrap iterations for both the bootstrap
schemes. We see that bootstrap variance is usually lower for the \emph
{subgraph subsampling} scheme as we increase the number of vertices of
the graph for different count statistics.
%

%

\subsection{Preferential attachment model}
\label{sec_pfa}
In the preferential attachment model, given $k$ initial vertices,
$k+1$th vertex attach to one of the preceding $k$ vertices with
probability proportional to degree. Now we have degree of vertex $v$,
defined as $D_v$ and $\bar{D} = \frac{1}{n}\sum_{v=1}D_{v}$. Also,
we have
\[
\tau(v) \simeq\frac{D_v}{\bar{D}}.
\]
Thus following equation \eqref{eq:nonpar_model}, we have the
probability of edge formation as
\[
w(u,v) = \frac{\tau(u)}{T(u)}\mathbf{1}(u\leq v) + \frac{\tau
(v)}{T'(u)}\mathbf{1}(v
\leq u),
\]
where $T(u) = \int_u^1 \tau(s)\,ds$ and $T'(v) = 1-T(v)$ and
\[
\tau(u) = \int_0^1 w(u, v)\,dv.
\]

Now the preferential attachment model can be defined by the following
formula on $w$:
\[
w(u, v) =  \frac{\tau(u)}{\int_u^1 \tau(s) \,ds} \mathbf{1}(u\leq v) + \frac{\tau(v)}{\int_v^1 \tau(s) \,ds}
\mathbf{1}(v\leq u).
\]

Thus for
\[
w(u, v)  =  (1-u)^{-1/2}(1-v)^{-1/2},
\]
we have
\[
\tau(v) = c(1-v)^{-1/2},
\]
which is equivalent to power law of \emph{degree distribution} $F
\equiv\tau^{-1}$.

%
\begin{figure}[b]
\centering
\begin{tabular}{@{}cc@{}}

\includegraphics{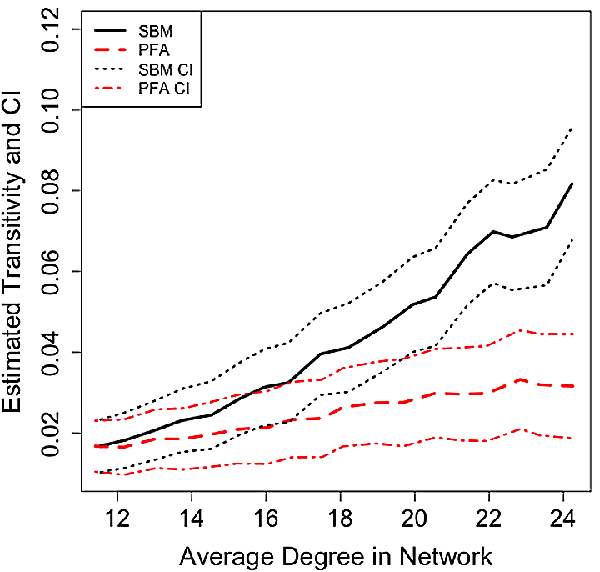}  & \includegraphics{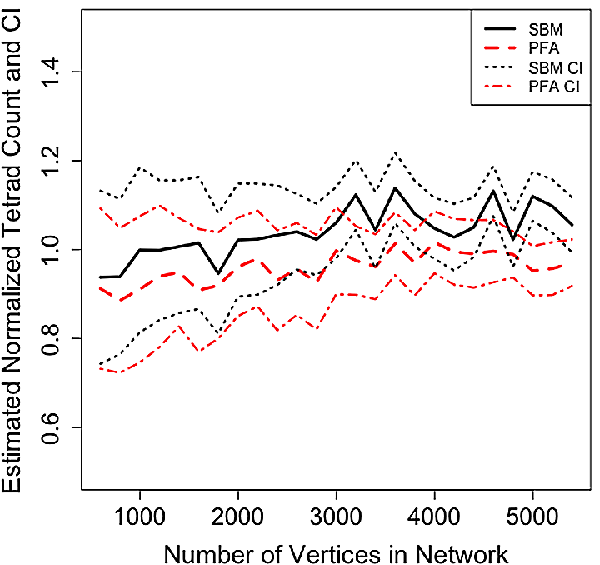}\\
\footnotesize{(a)} & \footnotesize{(b)}
\end{tabular}
\caption{\textup{(a)} For $n=1000$ we vary $\gl_n$, and we plot estimated
transitivity $\hat{T}^{B2}_{\mathrm{Tr}}$ and their 95\% confidence interval
(CI), where CI is estimated using bootstrap estimates of variance of
the estimators. \textup{(b)} We vary $n$, and we plot estimated normalized
tetrad count, $\hat{T}_{B2}(R)$, $R=$ tetrad and their 95\% confidence
interval (CI).}
\label{fig_sbmpfa}
\end{figure}

We simulate networks from both stochastic block models and preferential
attachment models, and then we try to compare the distribution of count
statistics of the graph for two different networks.
In Figure~\ref{fig_sbmpfa}(a) we vary the parameters of SBM as
$\mathbf{F} = \mu F^{(1)} + (1-\mu)F^{(2)}$, where $F^{(1)} = \operatorname{Diag}(0.035, 0.065)$ and $F^{(2)} = 0.001\mathbf{1}_2$. We increase
$\mu$ to increase $\gl_n$ and SBM have more pronounced \emph{cluster}
structure. We keep the average degree, $\gl_n$, of the two simulated
networks the same, and then we try to get the asymptotic distribution
of the transitivity statistic, $\hat{T}_{\mathrm{Tr}}$, for the two cases for
each $\gl_n$. We see here that for low $\gl_n$, we cannot
statistically distinguish between the transitivity of networks
generated from two different models, but they become statistically
distinguishable as average degree, $\gl_n$ and $\mu$, increase. In
Figure~\ref{fig_sbmpfa}(b), take SBM as in Section~\ref{sec_sbm} and PFA as in
Section~\ref{sec_pfa} keeping the average degree, $\gl_n$, of the two simulated
networks the same, and vary $n$, and we can statistically distinguish
the normalized tetrad count of networks between the two different
models for large $n$ based on \emph{subgraph subsampling} scheme.

%

\section{Real data examples}
\label{real data}
Social networks recently has become quite large after the introduction
of social networking sites. We consider two different social networks
as a platform for our experiments. The first one, high school romantic
relations data, is a small social network, whereas the second one,
Facebook college social network, has a greater number of nodes and
links. For both cases we use a \emph{subgraph subsampling} bootstrap scheme.

\subsection{High school network}
\label{highschool}
In this application, we try to quantitatively verify some of the
hypotheses mentioned by the authors of \cite{bearman2004chains} when
presenting the data. The network here is formed by students of
Jefferson High School as nodes, and if two students have romantic
relations, then there exists a link between those two nodes. In the
paper \cite{bearman2004chains} where the data was presented, an
observation was made about the dearth of short cycles in the network.
Our application here is trying to answer the question of whether the
absence of short cycles in this graph is significant or not. We
consider a very simple model for the data.

We consider that the data has been generated from two different models:
\begin{longlist}[(a)]
\item[(a)] Stochastic block model with two blocks (Male and Female),
and the connection probability matrix is given by
\[
P = \pmatrix{ P_{11}&P_{12}
\vspace*{2pt}\cr
P_{12}& P_{22}},
\]
where $P_{ab} = \mbox{ the average number of edges between blocks } a
\mbox{ and } b$ in the network, where $a,b=1,2$ are the two blocks with
male${} = {}$1 and female${} = {}$2. In this network, we have $P_{11} =0$, $P_{12}
= 0.0058$ and $P_{22} = 0.000025$. The probability of belonging to the
two blocks is $(0.497,0.503)$.
%
\item[(b)] Preferential attachment model with $\rho= \frac{\gl_n}{n}$,
where $\gl_n = $ the average degree of the network $=1.66$ and $n$ is
the number of nodes.
\end{longlist}
Now, for these two simple models, we can theoretically find the
normalized count of small cycles. Then we can perform a hypothesis test
to find out whether the number of small cycles we see in this network
is significantly small or not. For both models, we can find $\til
{P}(R)$, where $R = $ the cycles of size 3 and 4 based on the
parameters defined for models in (a) and (b) and using equation \eqref
{eq_norm_par}, and we shall call it $\til{P}_0(R)$. Also, for the
network, the unknown integral parameter for the subgraph $R$ is $\til
{P}(R)$. Formally, the hypothesis becomes
\[
H_0: \til{P}(R) = \til{P}_0(R)\quad \mbox{vs}\quad \til{P}(R)
< \til{P}_0(R)
\]
for each $R$ and for each model (a) and (b). We use the results of
Theorem~\ref{gen_thm} to form the asymptotic test. The results are
given in Table~\ref{tbl_hs1}.
We see in the results that according to the two simple models, it is
extremely unlikely for 3-cycles and 4-cycles to occur in the graph. In
fact, the original network has \textit{too many} 4-cycles short cycles,
not \textit{too few}. This is an interesting observation coming out of
our simple exploratory analysis. Thus our simple models do not capture
the probabilistic mechanism of the original network correctly, and we
need to analyze the short cycles in the network more closely to
understand their formation.
%
\begin{table}
\caption{The normalized subgraph counts, their standard deviation and
the expected counts from the stochastic block model (SBM) and
preferential attachment model (PFA) for the whole high school~network}
\label{tbl_hs1}
\begin{tabular*}{\textwidth}{@{\extracolsep{\fill}}lcccc@{}}
\hline
\textbf{Subgraph} & \textbf{Normalized count} & \textbf{Standard deviation} & \textbf{Count (SBM)} & \textbf{Count
(PFA)} \\
\hline
$(1,2)$-wheel & 2.27 & 0.17 & 1.01 & 2.97 \\
3-cycle & 1.31 & 0.1\phantom{0} & 0.01 & 1.04 \\
4-cycle & 9.47 & 3.16 & 0.63 & 3.06\\
\hline
\end{tabular*}
\end{table}


Note that this is a very small and sparse network. For this network,
the use of Theorem~1 from \cite{MR2906868} would have sufficed, but we
give the example as an example of the use of count statistics and their
quantitative behavior. In \cite{bearman2004chains}, simulation-based
tests were used.

Comparison of count statistics in the social network literature has
been based on parametric simulation \cite{MR2724362} or data bank
related tests \cite{wasserman1994social}. In these tests, the networks
are generated from either a random graph model or from a data bank of
networks (as in \cite{davis1967structure}). Permutation of nodes' block
identity-based tests are used for fitting block models \cite
{wasserman1994social}. We use asymptotically Gaussian tests based on
nonparametric exchangeable models for comparing graphs. The
hypothetical model we consider is nonparametric and thus more general
than simulation-based tests on specific random graph models or data
bank-based tests. Permutation of nodes' block identity-based tests seem
to function more as measures of goodness of fit of the block model
assignment of the particular graph.


\subsection{Facebook network}
\label{facebook}
In this application, we try to quantitatively analyze the behavior of
some of the known descriptive statistics for Facebook collegiate
networks. The networks were presented in the paper by Traud et al.
\cite
{MR2834086}. The network is formed by Facebook users acting as nodes,
and if two Facebook users are ``friends'' there is an edge between the
corresponding nodes. Along with the network structure, we also have the
data on covariates of the nodes. Each node has covariates: gender,
class year and data fields that represent (using anonymous numerical
identifiers) high school, major and dormitory residence. We try to
answer two very basic questions quantitatively for these networks:
\begin{longlist}[(1)]
\item[(1)] Can the node covariates act as cluster identifiers?
\item[(2)] Can two college networks be distinguishable in terms of some
basic descriptive statistics?
\end{longlist}
In order to address the first question, we consider the network of a
specific college (Caltech). We consider the covariates class year,
major and dormitory residence as our covariates of interest. We take
the induced network created by levels of each of these covariates and
try to see if those networks have different clustering properties. For
example, consider class year and major as the covariates of interest.
We consider the nodes belonging two different class years and find
their induced network from the whole collegiate network. Similarly, we
consider the nodes belonging two different majors and find their
induced network from the whole collegiate network.
Now, we have two different networks: one having nodes coming
exclusively from two different class years and the other having nodes
coming exclusively from two different majors. We now try to find which
of the two networks is more ``clustered'' by comparing the \textit
{transitivity} of the two networks. We can repeat the same exercise for
any two covariates and choose a subset of their levels. For the two
networks, the unknown integral parameter for transitivity is $\til
{P}^1_{\mathrm{Tr}}$ and $\til{P}^2_{\mathrm{Tr}}$, respectively. Formally, the
hypothesis becomes
\[
H_0: \til{P}^1_{\mathrm{Tr}} = \til{P}^2_{\mathrm{Tr}}\quad
\mbox{vs}\quad \til{P}^1_{\mathrm{Tr}} \neq \til{P}^2_{\mathrm{Tr}}.
\]

The second question can also be answered in a spirit similar to the
first. We consider the full collegiate network of two different
colleges (Caltech and Princeton). Then, we try to compare the
transitivity of these two collegiate networks. For the two networks,
the unknown integral parameter for transitivity is $\til{P}^1_{\mathrm{Tr}}$ and
$\til{P}^2_{\mathrm{Tr}}$, respectively. Formally, the hypothesis becomes
\[
H_0: \til{P}^1_{\mathrm{Tr}} = \til{P}^2_{\mathrm{Tr}}
\quad\mbox{vs}\quad \til{P}^1_{\mathrm{Tr}} \neq \til{P}^2_{\mathrm{Tr}}.
\]

These comparisons could, in principle, be possible using the results
given in Bickel et al. \cite{MR2906868}, but they are computationally
intractable. Using bootstrap estimators, we can estimate the variance
of the estimators and thus perform hypothesis testing in reasonable time.

In Tables~\ref{tbl_fb1}, \ref{tbl_fb2} and \ref{tbl_fb3}, we present an
excerpt of the result of our analysis and answer both of the questions.
These results give a better understanding about the network statistics
reported in \cite{MR2834086}, like those reported in Table~3.1 of
\cite
{MR2834086}. Using the numerical comparison of the transitivity values
reported in the table of \cite{MR2834086} alone can be statistically
unreliable, without a proper testing of whether the difference in
values for different networks is statistically significant. Such
comparison statements are now possible to make with the methods
proposed in this paper.

\begin{table}
\caption{Transitivity of induced networks formed by considering only
two levels of a specific covariate of a specific collegiate network}\label{tbl_fb1}
\begin{tabular*}{\textwidth}{@{\extracolsep{\fill}}lccc@{}}
\hline
& \textbf{Class year (CY)} & \textbf{Dormitory (DM)} & \textbf{Major (MJ)} \\
\hline
Estimated transitivity & 0.15 & 0.22 & 0.12 \\
\hline
\end{tabular*}
\end{table}
%
\begin{table}[b]
\tablewidth=230pt
\caption{The difference between class year and dorm is not significant,
but the difference between dorm and major is significant by an
asymptotic normal test at 5\% level. The data was presented in Traud et
al. \cite{MR2834086}}
\label{tbl_fb2}
\begin{tabular*}{230pt}{@{\extracolsep{\fill}}lcc@{}}
\hline
\textbf{Difference} & \textbf{CY and DM} & \textbf{DM and MJ} \\
\hline
Estimated & 0.07 & 0.1 \\
Estimated SD & 0.05 & 0.035 \\
\hline
\end{tabular*}
\end{table}

%
%
%

Now, without finding the bootstrap estimate of count statistics and its
variance, finding the asymptotic distribution of these count statistics
will not be possible. Thus with the help of the bootstrap-based estimates,
we can perform hypothesis testing on the count statistics and provide
the estimates of their asymptotic distribution.

\section{Conclusion and future works}
In this paper, we have considered two known subsampling schemes of
networks and have tried to show situations where they are applicable to
finding the asymptotic distribution of certain \textit{count
statistics} of the network.
We have showed that the normalized bootstrap subsample estimates of the
count statistics and their smooth functions have asymptotic normal
distribution. We have proposed bootstrap schemes by which we can
efficiently compute the asymptotic mean and variance of these count
statistics. We have also showed that the \textit{subgraph sampling}
bootstrap scheme seems most stable, and we recommend using this scheme
as bootstrap subsampling scheme in most cases.

\begin{table}
\tablewidth=230pt
\caption{The difference of transitivity between two networks is not
significant by an asymptotic normal test at 5\% level. Therefore
Network 1 cannot be said to be more \textit{``clusterable.''} The data
was presented in Traud et al. \cite{MR2834086}}
\label{tbl_fb3}
\begin{tabular*}{230pt}{@{\extracolsep{\fill}}lcc@{}}
\hline
& \textbf{Network 1} & \textbf{Network 2} \\
\hline
Estimated transitivity & 0.29 & 0.16 \\
Estimated difference & \multicolumn{2}{c}{0.13} \\
Estimated difference SD & \multicolumn{2}{c}{0.11} \\
\hline
\end{tabular*}
\end{table}

We also use the estimated asymptotic mean and variances of the count
statistics to construct hypothesis tests. These hypothesis tests can
serve several purposes, such as:
\begin{longlist}[(a)]
\item[(a)] distinguishing between the count statistics of two different
networks;
\item[(b)] distinguishing between parts of same network;
\item[(b)] testing whether a network has been generated from a specific
model by comparing the empirical and population versions of the count statistic;
\item[(c)] testing how close parameters of two different network models
can become.
\end{longlist}
All of these different qualitative tests can be made quantitative by
using hypothesis tests using the count statistics. Using subsample
bootstrap estimates of count statistics, we show from simulations that
transitivity of networks from stochastic block models becomes easier to
differentiate from transitivity of the preferential attachment model as
the average degree grows. Similarly, in real networks, such as the
Facebook collegiate network, we show that certain covariate-based
subnetworks have more ``cluster'' structure than others. Also, even in
large networks, conclusions based only on means, as opposed to
confidence statements using variances, could be unreliable.


\subsection{Future works}
One natural generalization could be the use of a bootstrap scheme to
get asymptotic distribution of global statistics, such as graph cut,
conductance, functionals of graphon (nonintegral functionals) and such
parameters. Sample and bootstrap estimates of such parameters are
sometimes obtainable, but their theoretical properties are still
unknown. It would be a nice future endeavor to extend our bootstrap
subsampling scheme to estimate such global characteristics of the networks.

\section*{Acknowledgments}
We thank Aiyou Chen, Dave Choi and Liza Levina for helpful discussions
and comments.


\begin{supplement}[id=suppA]
\stitle{Supplement to ``Subsampling bootstrap of count features of networks''}
\slink[doi]{10.1214/15-AOS1338SUPP} 
\sdatatype{.pdf}
\sfilename{aos1338\_supp.pdf}
\sdescription{In the Supplement, we prove Theorems~\ref{int_thm1},
\ref{int_thm2}, Proposition~\ref{prop_var_est}, Lemmas \ref
{lem_var_est1} and \ref{lem_boot_var}.}
\end{supplement}

\printaddresses
\end{document}